\documentclass[5p,times]{elsarticle}

\usepackage[T1]{fontenc}
\usepackage{textcomp}
\usepackage{amsmath,amssymb}
\usepackage{graphicx}
\graphicspath{{figures/}}
\usepackage{booktabs}
\usepackage{array}
\newcolumntype{L}[1]{>{\raggedright\arraybackslash}p{#1}}
\usepackage{algorithm}
\usepackage[noend]{algpseudocode}
\usepackage{xurl}
\usepackage[hidelinks]{hyperref}
\biboptions{sort&compress}
\newcommand{\sig}[1]{\texttt{#1}}
\newcommand{\lastseq}{\mathtt{last\_seq}}
\newcommand{\us}{\textmu s}

\journal{Microprocessors and Microsystems}

\begin{document}

\begin{frontmatter}

\title{WADJET: Whole-Job Authorization and Dispatch for Job Execution
Trust on FPGA System-on-Chip Accelerators}

\author[cpp,uiuc]{Mohamed El-Hadedy\corref{cor1}}
\ead{mealy@cpp.edu}
\ead{hadedy@illinois.edu}
\cortext[cor1]{Corresponding author.}
\affiliation[cpp]{organization={Department of Electrical and Computer
  Engineering, California State Polytechnic University, Pomona},
  addressline={3801 West Temple Avenue},
  city={Pomona}, state={CA}, postcode={91768}, country={USA}}
\affiliation[uiuc]{organization={Coordinated Science Laboratory,
  University of Illinois Urbana-Champaign},
  addressline={1308 West Main Street},
  city={Urbana}, state={IL}, postcode={61801}, country={USA}}

\begin{abstract}
On an FPGA system-on-chip, the software that submits work to an
accelerator also decides what the accelerator does. It writes the
registers one by one, and the hardware runs whatever it finds, so faulty
or compromised software can change one address or resend a job. Existing
protections decide who may reach an accelerator or keep its data safe.
Deciding whether one job was approved and is still unused takes three
things in one place: the only way to the accelerator, the whole job held
back until checked, and a record of used approvals. WADJET puts the
three in one gate in FPGA logic. A separate authority encrypts and
authenticates each approved job, with a sequence number, using
Ascon-AEAD128. The gate verifies the whole job before releasing any
part, accepts only a newer sequence number, uses it up, and writes the
registers of an unmodified accelerator itself. The gate replaces only
the processor's direct line to the registers and works beside the
existing protections. In a software model, a check on each write ran
jobs mixed from two approved ones; the whole-job rule released none. Its
release path is proven on the Verilog for all inputs. On a Zynq-7020
running a language model, it released all 21,000 approved jobs and none
of 14,700 altered or repeated ones. It needs a third of the lookup
tables of ShEF's register protection, which protects more, and adds
about ten microseconds to starting a job. It assumes trusted control of
its key, reset, and configuration.
\end{abstract}


\end{frontmatter}

\section{Introduction}
\label{sec:intro}

Field-programmable gate array (FPGA) system-on-chip (SoC) devices put
an application processor and programmable logic on one chip. Designers use the logic for
accelerators, and for embedded machine learning (ML) a common one is a
matrix-multiplication engine~\cite{sze2017efficient,xu2024llamaf}. The
processor controls the accelerator through a very simple interface. It
writes a few registers that say where the inputs are, where the weights
are, where the result should go, and how large the matrices are. Then
it sets a start bit, and the accelerator runs.

\begin{figure*}[!t]
\centering
\includegraphics[width=\textwidth]{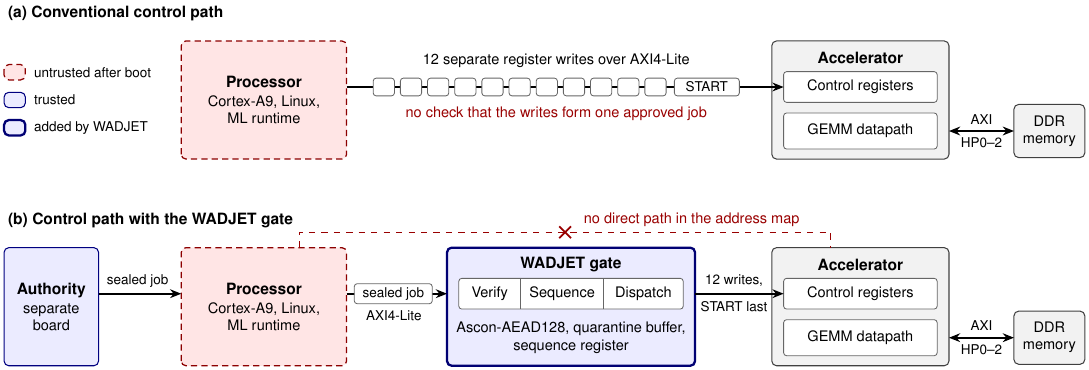}
\caption{Accelerator control path on an FPGA SoC: (a) conventional;
(b) with the WADJET gate.}
\label{fig:system}
\end{figure*}

This interface was designed to be convenient for software, and it does
not check what it is asked to do. Fig.~\ref{fig:system}(a) shows the
weakness. Each register write travels over the bus on its own and is
accepted on its own. The accelerator cannot tell whether the writes it
has received belong together, and it runs whatever is in its registers
when the start bit arrives. A job is therefore never a single unit in
hardware. It is assembled piece by piece, and any piece can be changed.

A single changed write is enough to cause harm. A different output
address sends the result to the wrong buffer, and a different size
makes the accelerator read or write beyond the intended data. An old
job sent again unchanged repeats work that was approved only once.
Whether the cause is a driver bug, a faulty update, or an attacker in
control of the processor's software, the hardware runs the job.

Embedded SoCs now run language models on such
accelerators~\cite{xu2024llamaf}, and one pass of the model used in
this work sends 210 jobs across this interface
(Section~\ref{sec:eval}). The software between an application and the
accelerator's registers has also grown, from the operating system and
drivers to the runtime and the model-serving code. In a recent
incident, an autonomous artificial intelligence (AI) agent gained access to production ML
infrastructure~\cite{huggingface2026intrusion,openai2026hfevalincident},
which shows that software running with valid credentials cannot be
assumed to behave as intended. Many embedded SoCs also use processors
such as the Arm Cortex-A9, whose only hardware security extension is
TrustZone, which splits the processor into a secure and a normal
world~\cite{pinto2019trustzone}.

Strong protections already surround accelerators, but none was built
for this fault. The software that changes one address or resends a job
is not an outsider. It is the accelerator's own operator, with every
right to use the bus. The open question therefore concerns the action
and not the party: was this particular job approved, and has that
approval already been used? Settling it takes three things in one
place. That place must be the processor's only way to the accelerator,
in hardware that software cannot change. The whole job must be held
back there until all of it has been checked. And approvals that have
been used must be remembered. No existing kind of protection offers all
three on this class of device (Section~\ref{sec:related}). Access
control is the only way in, but it judges one write at a time and does
not know what a job is. Secure channels check each message and pass it
on, so nothing is held back. Trusted monitors do judge a whole task,
but as software on the processor, and they need Arm's newer
architecture or a monitor in TrustZone, which has known weaknesses
where it meets programmable logic~\cite{benhani2019trustzone}. Checks
built into the device need a modified accelerator, whereas an
accelerator generated by high-level synthesis is used as it is.

WADJET puts the three in one place. A separate authority approves a job
by sealing it, that is, by encrypting and tagging the complete list of
register writes, together with a sequence number, under a key that the
software does not hold. A gate in the FPGA logic stands between the
processor and the accelerator's registers
(Fig.~\ref{fig:system}(b)), and the processor has no other path to
them. The gate holds the job until the whole of it is verified, checks
that it is newer than any job it has accepted, uses up the approval,
and then performs the register writes itself. If either check fails, no
write reaches the accelerator. Software still submits every job, but it
no longer decides what the accelerator does. The gate replaces only the
processor's direct line to the registers. It works beside the existing
protections: they keep deciding who may reach the accelerator and keep
its data safe, and the gate decides which job may start.

The gate invents no cryptography. It uses the standard authenticated
cipher Ascon-AEAD128 as it is~\cite{turan2025ascon}; the tag and the
rule that a nonce is never used twice come with such a cipher, as they
do with the authenticated ciphers of the systems reviewed. The
contribution is architectural. Two things exist separately in earlier
work: a check of a whole task before it starts, done by a software
monitor or inside a modified device, and a check in FPGA logic in front
of an unmodified accelerator, done one message at a time. WADJET does
both at once. It checks one complete job, approved once and used up at
the moment of release, in fixed logic in front of an unmodified
accelerator and with no software monitor on the processor. A check on
each write keeps the writes in order; only an approval of the whole job
keeps a half-delivered job out of the registers.

The rest of the paper follows the problem to its fix.
Section~\ref{sec:model} states the property, job-atomic release, and
the three conditions behind it. Section~\ref{sec:related} holds each kind of existing
protection against those conditions and states what a system gains when
the gate works beside it. Sections~\ref{sec:design}
and~\ref{sec:impl} build the gate and wire a Zynq-7000 SoC so that
nothing goes around it. Section~\ref{sec:eval} then shows that the fix
holds, by a proof on the gate's Verilog for all inputs and by tests on
hardware with a language-model workload, and what it costs against a
direct start. Section~\ref{sec:compare} sets that cost and the
guarantees beside ten published systems, and
Section~\ref{sec:discussion} discusses deployment and reuse.

\section{Problem and assumptions}
\label{sec:model}

This section fixes what a job is, what software may do to it, what is
trusted, and what the hardware must then guarantee. On the Zynq-7000
family, the processor reaches the programmable logic through
general-purpose ports (GP0 and GP1) of the Advanced eXtensible
Interface (AXI), and the logic reaches memory
through separate high-performance ports (HP0 to
HP3)~\cite{amd2026ug585}. An accelerator generated by high-level
synthesis (HLS) has two kinds of interface~\cite{amd2022ug1399}. A
small AXI4-Lite register block holds its arguments and its start and
done bits, and one AXI master port per data argument moves the bulk
data. AXI4-Lite is the simplest AXI bus, in which every write is a
separate exchange of one address, one 32-bit word, and one
response~\cite{arm2021axi}. The accelerator used in this work computes
one general matrix multiplication (GEMM) per job, and its job is the
set
\begin{equation}
J=\{\mathrm{ACT},\mathrm{WGT},\mathrm{OUT},M,N,K,\mathrm{CTRL},\mathrm{START}\},
\label{eq:job}
\end{equation}
where $\mathrm{ACT}$, $\mathrm{WGT}$, and $\mathrm{OUT}$ are the
addresses of the input activations, the weights, and the output; $M$,
$N$, and $K$ give the matrix sizes; $\mathrm{CTRL}$ holds control
settings; and $\mathrm{START}$ launches the run. On the bus, $J$ is a
fixed list of 12 register writes: two for each of the three addresses,
one for each size, two for the control settings, which enable the
accelerator's done interrupt, and the start bit
last. Each write is stored as a 64-bit record, with the register
address in bits 63--32 and the data in bits 31--0, so one job is 96
bytes.

The software that issues these writes is assumed to be untrustworthy.
After the system has booted and been set up, all software outside the
processor's secure world may be faulty or hostile, including the
operating system, the drivers, and the application. Such software can
issue any bus transaction to the programmable logic, and it can read,
drop, change, reorder, or resend any job that passes through it. This
software is the accelerator's own operator and not an intruder. The
question is not whether it may reach the accelerator, but whether each
job it submits was approved.

Four elements are trusted. The first is the \emph{authority}, a
separate board that decides which jobs are allowed. The second is the
gate's own logic and the FPGA configuration that contains it. The third
is the \emph{setup step}, in which trusted boot software loads a secret
key into the gate and locks it before any untrusted software runs.
Secure boot~\cite{amd2026ug585} and the secrecy of that key are taken
as given. The fourth is control of the programmable logic, meaning its
reset, its clock, and its configuration. A reset clears the gate's key,
its key lock, and its memory of past jobs, so untrusted software must
not be able to reset or reconfigure the logic and load a different key.
This fourth element is a requirement on the platform and not something
the gate enforces, and the results of Section~\ref{sec:eval} hold
under it (Section~\ref{sec:discussion}).

Physical attacks, side channels, fault injection, and denial of service
are outside the scope, and so is the protection of data stored in
memory. Fixing the addresses in a job fixes \emph{where} the
accelerator reads and writes, but it does not stop the data stored
there from being changed. The gate's guarantee concerns the jobs.
Whether the data at those addresses is the intended data is a separate
guarantee, which the data-protection systems of
Section~\ref{sec:related} supply, and WADJET can work beside them.

Under these assumptions, the hardware should provide one property:
\emph{the accelerator's registers are written only as part of a job
that the authority issued, with that job's writes in the authority's
order; no job is released after a later one, and none is released
twice.} This paper calls the property \emph{job-atomic release}: the
decision to release is taken once for the whole job, before any of its
writes, and at most once. Software can still withhold a job. Later jobs are then
released, and the withheld one can no longer run. Three conditions make
this property hold. The gate must be the only bus path from the
processor to the accelerator's registers. No part of a job may reach
the accelerator before the whole job has been verified. And the gate
must mark a job as used before it starts releasing it.
Section~\ref{sec:related} holds the published protections against
these three conditions. Section~\ref{sec:design} then provides the
second and third inside the gate, and Section~\ref{sec:impl} provides
the first in the wiring around it.

\section{Related work}
\label{sec:related}

\begin{figure*}[!t]
\centering
\includegraphics[width=\textwidth]{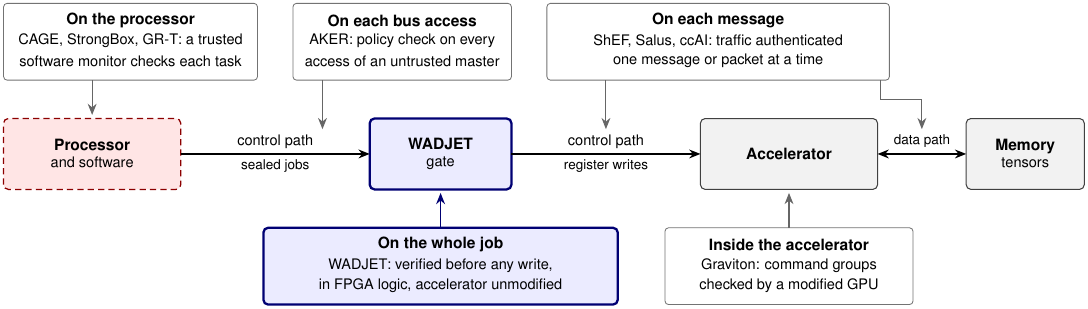}
\caption{Enforcement points of prior work and of WADJET.}
\label{fig:position}
\end{figure*}

Each kind of published protection solves a real problem, and this
section asks two things of each: which of the three conditions of
Section~\ref{sec:model} it meets, and what a system gains when the gate
works beside it. Most of the designs build a trusted execution
environment (TEE), an isolated environment, also called an enclave,
that stays protected when the operating system is compromised. A recent
survey sorts 51 accelerator-TEE studies by where access is
controlled~\cite{wang2026sok}, and a systematization of 27 TEEs built
on SoC FPGAs finds that most serve one feature or
application~\cite{perkins2025sok}. Fig.~\ref{fig:position} groups the
designs closest to this work by the place of the check, and
Table~\ref{tab:features} lists what the main ones guarantee.

\begin{table*}[!t]
\centering
\caption{Comparison of guarantees with published systems (\checkmark:
yes; $\circ$: partly; --: no).}
\label{tab:features}
\small
\setlength{\tabcolsep}{4pt}
\begin{tabular}{@{}L{0.37\textwidth} c c c c c c c c@{}}
\toprule
\textbf{Guarantee} & \textbf{AKER} & \textbf{ShEF} & \textbf{Salus} & \textbf{ccAI} & \textbf{CAGE} & \textbf{StrongBox} & \textbf{Graviton} & \textbf{WADJET}\\
\midrule
Control writes or commands are authenticated & -- & \checkmark & \checkmark & \checkmark & \checkmark & $\circ^{b}$ & \checkmark & \checkmark\\
\addlinespace
A whole job or task is verified before it starts & -- & -- & -- & -- & \checkmark & \checkmark & \checkmark & \checkmark\\
\addlinespace
Repeated messages or jobs are rejected & -- & $\circ^{c}$ & \checkmark & \checkmark & $\circ^{d}$ & $\circ^{e}$ & \checkmark & \checkmark\\
\addlinespace
The check runs in FPGA logic, with no CPU enclave, software monitor, or modified device & \checkmark & \checkmark & $\circ^{f}$ & $\circ^{f}$ & -- & -- & -- & \checkmark$^{a}$\\
\addlinespace
The accelerator's data is protected & $\circ^{g}$ & \checkmark & $\circ^{h}$ & \checkmark & \checkmark & \checkmark & \checkmark & --\\
\bottomrule
\multicolumn{9}{@{}p{0.98\textwidth}@{}}{\footnotesize $^{a}$Trusted
software loads the key at start-up and guards the reset and
configuration of the logic; it checks no jobs. $^{b}$Checks the
registers that the untrusted driver wrote but does not authenticate
them. $^{c}$No counter is described for register messages. $^{d}$The
signature covers the expected order of tasks; version numbers against
rollback are proposed, not built. $^{e}$Detects a changed order of
tasks; protection against a repeated task is not described. $^{f}$The
check runs in FPGA logic but relies on an enclave or a confidential
virtual machine on the host. $^{g}$Isolates memory regions without
encrypting them. $^{h}$Data encryption is left to the accelerator's
developer.}
\end{tabular}
\end{table*}

Access control guards the path, which is the first condition, and
stops there. AKER wraps
the untrusted masters of an SoC with policy checkers that decide
whether each individual access is allowed~\cite{restuccia2021aker}, and
earlier hardware firewalls for FPGA-based systems follow the same idea
with added memory encryption~\cite{cotret2016firewalls}. Self-secured
devices move the TrustZone check into each peripheral built in
Zynq-7000 logic, which filters every access by the bus protection
signals~\cite{pinto2021selfsecured}. CapChecker bounds every memory
access of an accelerator with hardware capabilities and needs a
processor with CHERI (Capability Hardware Enhanced RISC
Instructions)~\cite{cheng2025capchecker}. These checks are cheap and
fast, and they keep strangers out. But a rule about single accesses
holds nothing back and remembers nothing, so it cannot tell whether a
group of allowed writes forms an approved job or whether that job was
sent before. Beside such a check, the gate covers the case it leaves
open, in which the allowed software itself goes wrong: reaching the
registers is no longer enough to start a job.

Secure channels protect a user's data from the platform and judge one
message at a time. ShEF surrounds a cloud-FPGA accelerator with a
Shield that encrypts and authenticates its register and memory traffic
for a remote user~\cite{zhao2022shef}. SGX-FPGA pairs an enclave of
Intel Software Guard Extensions (SGX) with a secure monitor in the FPGA
and encrypts the traffic between them~\cite{xia2021sgxfpga}. Salus
builds a TEE across a central processing unit (CPU) and an FPGA, with a
session counter on register transactions, and assumes a TEE-enabled
host CPU, which is SGX in its prototype~\cite{zou2024salus}. ccAI
places a security controller built in FPGA logic between the host and
an unmodified accelerator, and it filters and integrity-checks every
packet of the PCI Express bus on behalf of a confidential virtual
machine~\cite{wang2025ccai}. ShEF, Salus, and ccAI work in FPGA logic,
as WADJET does, but each register write or packet is checked as it
arrives and passed on. The second condition is therefore not met: part
of a job can be in the accelerator before a later write fails. What
these designs give is what the gate lacks, namely protection of a
remote user's data and, in ShEF and Salus, attestation of the design. A system
that pairs a channel of this kind with the gate has both: the channel
keeps the data safe, and the gate lets a job reach the registers only
whole and only once.

Trusted software on the processor comes closest in what it checks.
ACAI extends Arm's Confidential Compute Architecture (CCA), an Armv9
feature that isolates workloads under a small trusted
monitor~\cite{li2022cca}, to accelerators~\cite{sridhara2024acai}, and
HIX places the driver of a graphics processing unit (GPU) in an
enclave~\cite{jang2019hix}. On a Zynq UltraScale+ device, T-Edge maps
the accelerator's interface to the TrustZone secure world, so that only
trusted software can write its registers or reconfigure the
logic~\cite{shen2024tedge}, and FPGA-TrustZone extends the protection
of TrustZone to accelerators in the logic of an FPGA
SoC~\cite{wang2025fpgatrustzone}. Three designs reason about a whole
task. In CAGE, a trusted monitor in CCA verifies each accelerator task
and writes the start command itself~\cite{wang2026cage}. StrongBox does
the same for a GPU on Arm devices with a monitor in
TrustZone~\cite{deng2022strongbox}. GR-T records the register accesses
of a GPU workload in the cloud, signs the recording, and lets a small
replayer in TrustZone verify and replay it~\cite{park2023grt}; such a
recording is meant to be replayed many times. These designs hold a
whole task back, as the second condition asks, but the place that does
so is software on the processor, which handles every task and must
itself stay correct. WADJET makes the same decision in fixed logic with
no monitor on the processor; Section~\ref{sec:discussion} sets out what
it asks of trusted software. Beside such a monitor, the gate adds a
check that does not depend on the monitor's code: a job that the
authority did not approve is not released even when the software above
it is wrong.

A check inside the accelerator can meet all three conditions, but only
in a device that was changed to hold it. Graviton protects groups of
GPU commands with authenticated encryption and a counter against
dropped, repeated, or reordered commands. A modified GPU command
processor verifies them, and the application's copy of the session key
is meant to be kept in a CPU enclave; the prototype emulates the GPU
changes in software~\cite{volos2018graviton}. A patent describes
command buffers that are tagged in a host TEE and checked by the GPU
against a forward-only counter~\cite{pappachan2022patent}. Graviton is
the closest to this work in mechanism. Where a device can be changed,
the check can live inside it; where it cannot, as with an accelerator
generated by HLS, the gate supplies the same check from outside.

Other designs protect a whole workload and not a single start. BYOTee
moves the computation into soft-processor enclaves built in Zynq-7000
logic, with the Arm side untrusted~\cite{armanuzzaman2024byotee}.
GuardNN, prototyped on an FPGA, signs hashes of a run's input, weights,
output, and instruction sequence, so tampering is detected after the
run and not prevented before it~\cite{hua2022guardnn}. LGT4CG pursues
lightweight trusted execution for cloud GPUs~\cite{yu2026lgt4cg}, and
HETEE and Composite Enclaves extend trusted execution across several
devices~\cite{zhu2020hetee,schneider2021composite}. The aim here is
narrower, namely to decide whether one job may start.

The remaining mechanisms each hold one piece. gVirt inspects GPU
command buffers as guests submit them~\cite{tian2014gvirt}, and
Honeycomb validates GPU programs ahead of
time~\cite{mai2023honeycomb}; both rely on trusted software in the
submission path. An input--output memory management unit (IOMMU) limits
where a device may access memory, but it is configured by the operating
system~\cite{markettos2019thunderclap}, which is the software that is
not trusted here. sIOPMP performs this check in hardware configured by
a small secure monitor~\cite{feng2024siopmp}, but it bounds memory
accesses and does not judge the commands a device receives. Memoir
shows how to keep forward-only state, on which the third condition
rests, safe across restarts~\cite{parno2011memoir}. An
Ascon core on an Artix-7 FPGA comes nearest among them: it withholds
plaintext until the tag is verified and rejects repeated nonces with a
Bloom filter in block RAM~\cite{kurian2025ascon}. It meets the second
and third conditions for a message stream, but it stands in front of no
device and writes no registers. WADJET keeps its forward-only state in
one register and uses it to gate the register writes of an accelerator.

Table~\ref{tab:features} gives the outcome. The designs that
authenticate control traffic in FPGA logic, ShEF, Salus, and ccAI, do
not hold a job back until all of it is verified. The designs that do
verify a whole task before it starts, CAGE, StrongBox, and Graviton,
run the check in a software monitor on the processor or inside a
modified device. WADJET is the only system in the table that verifies a
whole job in FPGA logic, with no monitor on the processor and an
unmodified accelerator. The notes under the table give the reason for
each partial mark. The last row shows the division of work: the other
systems protect data, and WADJET leaves that task to them. The gate is
therefore the piece that the others lack and not a replacement for any
of them. The next three sections
show how the gate meets the three conditions, and
Section~\ref{sec:compare} then compares cost.

\section{Design of the gate}
\label{sec:design}

\begin{figure}[!t]
\centering
\includegraphics[width=\columnwidth]{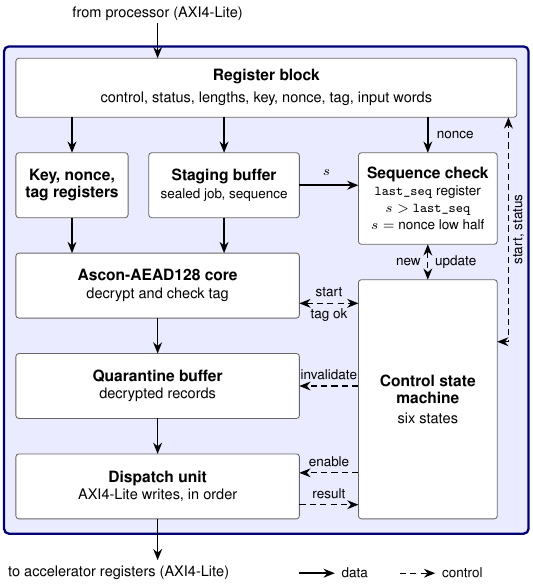}
\caption{Microarchitecture of the gate.}
\label{fig:gate}
\end{figure}

\begin{figure*}[!t]
\centering
\includegraphics[width=\textwidth]{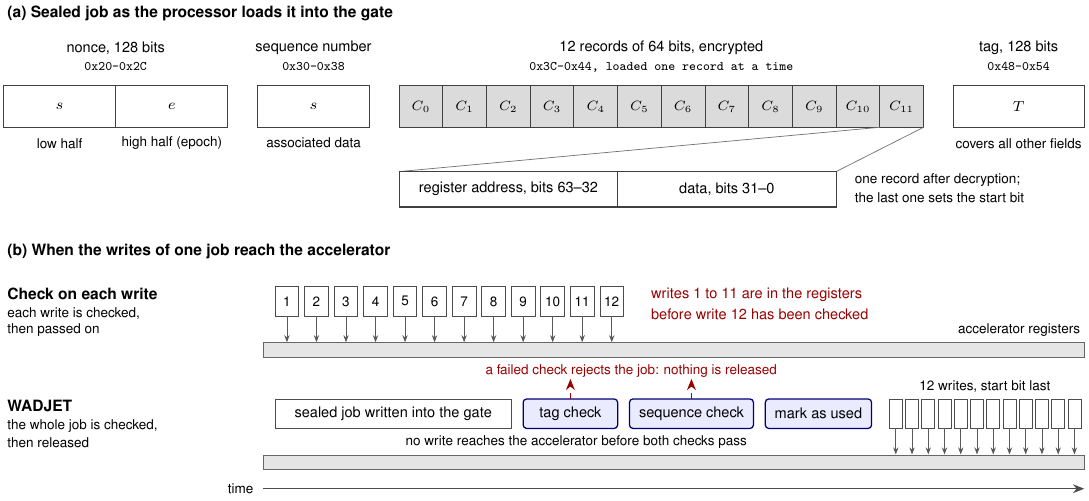}
\caption{Sealed job (a) and the time at which its writes reach the
accelerator (b).}
\label{fig:job}
\end{figure*}

The gate must supply the second and third conditions of
Section~\ref{sec:model}: nothing reaches the accelerator before the
whole job is verified, and a job is marked as used before it is
released. Fig.~\ref{fig:system}(b) shows where the gate sits,
Fig.~\ref{fig:gate} shows what is inside it, and
Fig.~\ref{fig:job}(a) shows the sealed job that it receives. A job
passes through four steps. The authority seals the 96-byte job with a
secret key that it shares with the gate, and attaches a sequence
number. The processor receives the sealed job and writes it into the
gate's input registers. The gate opens the job, verifies it, and checks
that its sequence number is new. Finally, the gate writes the 12
register values to the accelerator, with $\mathrm{START}$ last.
Algorithm~\ref{alg:dispatch} gives the procedure.

Jobs are sealed with Ascon-AEAD128, the lightweight cipher for
authenticated encryption with associated data (AEAD) standardized by
the National Institute of Standards and Technology
(NIST)~\cite{turan2025ascon,dobraunig2021ascon}. Ascon has
compact hardware implementations~\cite{gross2017ascon}, which suits a
gate that has to share a device with a large accelerator. The authority
encrypts the 96-byte job under a 128-bit key into 12 sealed records,
$C_0$ to $C_{11}$, and produces a 128-bit tag $T$, a check value that
only a holder of the key can compute. The
cipher also needs a 128-bit nonce, a value that must never be used
twice with one key~\cite{turan2025ascon}. Its low half is the 64-bit
sequence number $s$, and its high half is a 64-bit session number $e$,
called the \emph{epoch}; a session is the time between two trusted
restarts. The sequence number is also passed as associated data, which
is not encrypted but is covered by the tag. If any register address,
value, or position in the job is changed, the tag no longer matches,
and the same holds if the sequence number is changed. Encryption hides
the register values while the job passes through the processor. The
tag is what stops software from editing a captured job.

A cipher such as Ascon produces decrypted data before it has finished
checking the tag, and acting on such data is a known hazard, the
release of unverified plaintext~\cite{andreeva2014rup}. If the data
went straight to the accelerator, a forged job could configure it
before the check failed. The gate therefore separates three stages
(Fig.~\ref{fig:gate}). The \emph{staging} buffer holds the sealed job
that the processor wrote. The \emph{quarantine} buffer receives the
decrypted records ($P$ in Algorithm~\ref{alg:dispatch}). The \emph{dispatch} unit, which writes to the
accelerator, reads only from quarantine and is enabled only after the
tag check has passed, and there is no direct wire from the cipher's
output to the accelerator. While
a job is being processed, the gate ignores writes to its key, nonce,
length, sequence, job, and tag registers and accepts only a request to
clear its status flags. Software therefore cannot swap the job after
setting the gate's \sig{START} bit.

\begin{algorithm}[!t]
\caption{Handling of one sealed job by the gate.}
\label{alg:dispatch}
\small
\begin{algorithmic}[1]
\Require nonce $(s_n,e)$, sequence number $s$, sealed job $C$, tag $T$
\If{key not locked $\lor$ request not well formed}
  \State \textbf{reject} (\sig{CONFIG\_ERROR}); \Return
\EndIf
\State open $C$ with key, nonce, $s$, $T$; put records $P$ in quarantine
\If{tag does not match}\label{alg:tag}
  \State invalidate quarantine; \textbf{reject} (\sig{AUTH\_FAIL}); \Return
\EndIf
\If{$s \neq s_n \lor s \le \lastseq$}\label{alg:seq}
  \State invalidate quarantine; \textbf{reject} (\sig{FRESHNESS\_FAIL}); \Return
\EndIf
\State $\lastseq \gets s$ \Comment{used up before any write}\label{alg:update}
\If{record count does not match the declared length}
  \State invalidate quarantine; \textbf{reject} (\sig{CONFIG\_ERROR}); \Return
\EndIf
\For{each record $P_i$ in order}\label{alg:loop}
  \State write $P_i$'s data to $P_i$'s register address
  \If{the write returns a bus error}
    \State invalidate quarantine; flag \sig{AXI\_ERROR}; \Return
  \EndIf
\EndFor
\State flag \sig{DONE}\label{alg:done}
\end{algorithmic}
\end{algorithm}

To release each job at most once, the gate keeps one 64-bit sequence
register, $\lastseq$, which holds the highest sequence number it has
accepted since reset. The sequence number reaches the gate twice,
inside the nonce ($s_n$ in Algorithm~\ref{alg:dispatch}) and as
associated data ($s$). A verified job is released only if the two agree
and $s$ is higher than $\lastseq$ (line~\ref{alg:seq}), so no nonce is
accepted twice under one key. The gate updates $\lastseq$ \emph{before}
it releases the first write to the accelerator
(line~\ref{alg:update}). Because the update comes first, the same job
or any older one is rejected from then on, even if its tag is valid,
and a job that fails partway through cannot be sent again. The register
is zero after reset, so sequence numbers start at 1.

One register is enough to make every approval single-use, and the gate
does not require consecutive numbers. If software presents jobs out of
order, the newer one is released and the older one is then rejected. If
it withholds a job, that job never runs. Software can thus prevent
approved work, but it cannot run a job twice or out of order. Systems
that must accept jobs out of order would need a window of sequence
numbers instead, as network protocols use~\cite{kent2005rfc4303}.
Fig.~\ref{fig:job}(b) shows the result on a time line. Under a check on
each write, the first writes are in the accelerator before the last one
has been checked. Under WADJET, no write reaches the accelerator until
the whole job has passed both checks and its number has been marked as
used.

The sequence register covers one session, and sessions themselves are
separated by the key. A trusted restart resets the gate and loads a
fresh key and a new, nonzero epoch, so a job sealed in an earlier
session fails the tag check in a later one. The epoch safeguards the
cipher alone: should a key ever be loaded twice, the two sessions would
still use different nonces. It plays no part in rejecting old jobs, and
the gate neither stores nor compares it. Section~\ref{sec:eval} tests
both cases, a job repeated within a session and a job kept from an
earlier one.

The dispatch unit sends one register write at a time and waits for the
response before sending the next (line~\ref{alg:loop}), and it stops at
the first bus error. Because the authority always places
$\mathrm{START}$ last, a failure on any earlier write leaves the
accelerator configured but not running, and the used-up sequence number
prevents a retry of the same job.

The gate does not need to understand what the registers mean, since it
only passes on address--data pairs. The authority decides which
registers a job may write, and the bus address map limits the gate's
reach to the accelerator's register block. In its default
configuration, the gate accepts jobs of up to 32 records, the size of
its two buffers, of which the job format used here needs 12. It accepts the next job as soon as the
last write of the current one has been acknowledged and does not wait
for the accelerator to finish. Release and completion are therefore
different events. Release means that the gate has issued a job's
writes, which is what the property of Section~\ref{sec:model}
concerns; completion is the accelerator's own done signal, which the
gate does not observe. In the
experiments, the software waited for the accelerator's done interrupt
before it submitted the next job; a gate that itself waits for the
accelerator's idle signal is left for future work.

\section{Implementation}
\label{sec:impl}

The system is built on a Xilinx Zynq-7020 (xc7z020, clg400 package,
speed grade $-1$)~\cite{xilinx2018ds190} on a PYNQ-Z2
board~\cite{tul2019pynqz2}, with Vivado 2022.2, and all logic runs from
one 125-MHz clock. The accelerator is a Vitis HLS GEMM
kernel~\cite{amd2022ug1399} with one AXI4-Lite register block and three
AXI master ports for activations, weights, and outputs. The gate is
written in Verilog at the register-transfer level, around an Ascon
core set to decrypt and verify, and SmartConnect blocks route the
buses~\cite{amd2026pg247}. A software model of the cipher that
reproduces the NIST test vectors for
Ascon-AEAD128~\cite{turan2025ascon} seals jobs that the gate accepts,
which ties the hardware core to the standard for this job shape.
The gate's slave port has 32-bit data and an 8-bit address. Its master
port has 32-bit address and data and no read channel, so the gate
writes the accelerator's registers and never reads them.

The first condition of Section~\ref{sec:model} requires that software
have no other route to the accelerator's registers, and in this design
the bus address map provides none. Table~\ref{tab:addrmap} lists the
map, taken from the hardware description that Vivado generates with the
tested bitstream. From the processor, exactly one region of the
programmable logic is visible, which is the gate's register block, and
the processor's second general-purpose port is disabled. The
accelerator's registers appear only in the address space of the gate's
own AXI master, and no processor-side address leads to them. The
absence of a bypass therefore follows from how the buses are wired and
not from a software rule, and it holds as long as the loaded
configuration stays in place. An inspection of the deployed software
also found a single place where accelerator register writes are built,
and that place targets the gate. The accelerator's three data ports
keep their direct paths to memory, so bulk data never passes through
the gate, and each port is limited to its own 128-MB region.

The authority of the tested system runs on a separate Zynq UltraScale+
board, which the Zynq-7020 board reaches over a TCP connection. It
holds the key and the sequence counter. The software on the Zynq-7020
holds neither: it attaches to a gate whose key was loaded and locked in
the setup step, and it does not load a configuration or a key itself.
In the prototype, a provisioning program on the board carries out that
step before the job software starts: it loads the configuration, writes
the key that the authority supplies, locks it, reads the lock bit back
from the gate, and reports the result. This program and the channel
that carries the key to it are trusted in the prototype; they stand in
for the trusted boot software that Section~\ref{sec:model} assumes, and
the lock that the authority records is the one this program reports. At setup, the board also
registers the addresses of its three data buffers with the authority,
and the registration is then frozen. From then on, the software does
not choose what is sealed. The authority builds the 12 writes itself,
from the registered addresses and from matrix sizes that it holds, and
it issues a job only for a board whose provisioning record shows a
locked key. Every job of a board therefore carries the same three
addresses, and its sizes come from the authority and not from the
software. In this system, the gate fixes where the accelerator reads
and writes, how much, and which jobs may be started; the software places
each job's activations and weights in the buffers, and that data is
what Section~\ref{sec:model} leaves outside the gate's guarantee. Each
command reaches the board with two sealed packets, one for the job and
one for the single write that clears the accelerator's interrupt, under
two consecutive sequence numbers. The software stages the packets it is
given and refuses a command that arrives without them. Packets that are
withheld or delayed keep jobs from running, which is denial of service
and outside the scope. The time the authority needs to seal and deliver
a job was not measured.

\begin{table}[!t]
\centering
\caption{Bus address map of the tested design.}
\label{tab:addrmap}
\small
\setlength{\tabcolsep}{4pt}
\begin{tabular}{@{}llll@{}}
\toprule
\textbf{Master} & \textbf{Reaches} & \textbf{Base} & \textbf{Size}\\
\midrule
Processor (GP0) & Gate registers & \sig{0x4000\_0000} & 4 KB\\
Processor (GP0) & Accel.\ registers & not mapped & --\\
Gate master & Accel.\ registers & own space & 64 KB\\
Accel.\ ACT port & Memory (HP0) & \sig{0x0000\_0000} & 128 MB\\
Accel.\ WGT port & Memory (HP1) & \sig{0x0800\_0000} & 128 MB\\
Accel.\ OUT port & Memory (HP2) & \sig{0x1000\_0000} & 128 MB\\
\bottomrule
\end{tabular}
\end{table}

Table~\ref{tab:regs} lists the gate's registers as software sees them.
Once the key-lock bit is set, the key cannot be changed until the gate
is reset, and the gate rejects a start request unless the key is
locked. To submit a job, the
processor clears the two staged-word counters, writes the two lengths,
the nonce, the sequence number, the sealed records, and the tag, and
then sets the gate's \sig{START} bit. Every 64-bit word is written as
two 32-bit halves and is then appended to the staging buffer by a write
to its load register. One 12-record job takes 51 register writes in
this way, 50 to stage it and one to set \sig{START}, against 12 for an
unprotected start; the extra writes on the
processor side are the price of a boundary that the logic enforces. A request is well formed
when the key is locked, the sequence number is one 8-byte word, the job
length is a nonzero multiple of 8 bytes that fits the buffer, and the
loaded word counts match both lengths. The status register reports the
outcome as done (\sig{DONE}), authentication failure at the tag check
(\sig{AUTH\_FAIL}), freshness failure at the sequence check
(\sig{FRESHNESS\_FAIL}), bus error (\sig{AXI\_ERROR}), or
configuration error (\sig{CONFIG\_ERROR}), and these flags are cleared
at every start.

\begin{table}[!t]
\centering
\caption{Register map of the gate (32-bit AXI4-Lite).}
\label{tab:regs}
\small
\setlength{\tabcolsep}{3pt}
\begin{tabular}{@{}lL{0.72\columnwidth}@{}}
\toprule
\textbf{Offset} & \textbf{Purpose}\\
\midrule
\sig{0x00} & Control (write): bit 0 start; 1 clear staged-word counters; 2 clear status; 3 lock key (one-way)\\
\sig{0x04} & Status (read): bit 0 active; 1 cipher busy; 2 done; 3 authentication OK; 4 authentication fail; 5 bus error; 6 configuration error; 7, 8 buffer overflow; 9 key locked; 10 dispatch busy; 11 freshness fail\\
\sig{0x08}/\sig{0x0C} & Lengths in bytes: sequence number (must be 8); sealed job (multiple of 8, at most 256)\\
\sig{0x10}--\sig{0x1C} & 128-bit key (write-only)\\
\sig{0x20}--\sig{0x2C} & 128-bit nonce: sequence number (low half), epoch (high half)\\
\sig{0x30}--\sig{0x38} & Sequence number: 64-bit word in two halves, then a load strobe\\
\sig{0x3C}--\sig{0x44} & Sealed job: each 64-bit word in two halves, then a load strobe\\
\sig{0x48}--\sig{0x54} & 128-bit tag\\
\sig{0x58}/\sig{0x5C} & Word counts; version\\
\bottomrule
\end{tabular}
\end{table}

\begin{figure*}[!t]
\centering
\includegraphics[width=\textwidth]{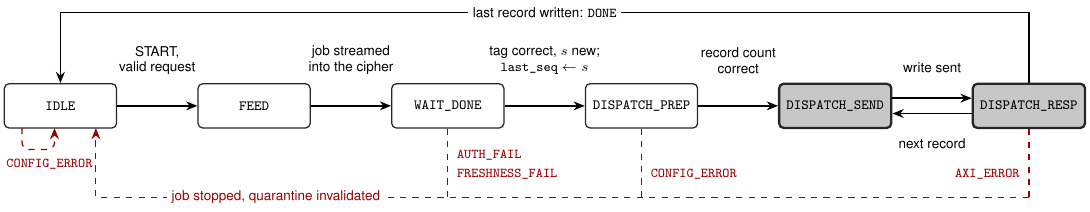}
\caption{Control state machine of the gate; the two dark states write
to the accelerator.}
\label{fig:fsm}
\end{figure*}

A six-state machine carries out Algorithm~\ref{alg:dispatch}
(Fig.~\ref{fig:fsm}). In \sig{IDLE} it checks that the request is well
formed and stays there with a configuration error if it is not. In
\sig{FEED} it streams the key, nonce, sequence number, sealed job, and
tag into the Ascon core, and in \sig{WAIT\_DONE} it waits for the tag
check. If the tag is wrong, the quarantine is invalidated and the
machine returns to \sig{IDLE}. If the tag is right, the machine checks
the sequence number and then either rejects the job or moves $\lastseq$
forward. \sig{DISPATCH\_PREP} confirms that the number of decrypted
records matches the declared length, and \sig{DISPATCH\_SEND} and
\sig{DISPATCH\_RESP} then write the records to the accelerator one by
one. These three states are the dispatch unit of Fig.~\ref{fig:gate}. A
quarantine overflow, which a well-formed request cannot cause, would
also end the job with a configuration error. No path reaches the two
writing states without passing both checks.

The routed design meets timing at 125~MHz with positive slack on setup
and hold, and the worst setup and hold paths in the timing report lie
in the accelerator. Table~\ref{tab:impl} gives resources and timing;
its lower rows are measurements discussed in Section~\ref{sec:eval}.
The gate, which contains the cipher core, both buffers, and both bus
ports, takes about 7\% of the device's lookup tables (LUTs), roughly three quarters
of what the accelerator uses, and 6\% of its flip-flops (FFs). It needs
no block RAM (BRAM) and no digital signal processing (DSP) blocks, because its buffers are built
from ordinary logic, and on this device the accelerator already
occupies every block RAM.

\begin{table}[!t]
\centering
\caption{Implementation results on the Zynq-7020 (post-route).}
\label{tab:impl}
\small
\setlength{\tabcolsep}{4pt}
\begin{tabular}{@{}lr@{}}
\toprule
\textbf{Quantity} & \textbf{Value}\\
\midrule
Gate LUTs (share of device) & 3,965 (7.45\%)\\
Gate FFs (share of device) & 6,685 (6.28\%)\\
Gate BRAM, DSP blocks & 0, 0\\
Accelerator LUTs, FFs (share of device) & 10.0\%, 7.8\%\\
Accelerator BRAM, DSP blocks & 140 (100\%), 132 (60\%)\\
Clock & 125 MHz\\
Setup, hold slack & $+0.048$, $+0.012$ ns\\
\addlinespace
\multicolumn{2}{@{}l}{\emph{Compiled driver, one board}}\\
Direct start, 12 register writes & 3.1 \us\\
Sealed job, 50 staging writes & 9.4 \us\\
Gate, START to DONE: median, maximum & 3.6, 4.3 \us\\
\addlinespace
\multicolumn{2}{@{}l}{\emph{Python runtime of the campaign}}\\
START to outcome, median & 59.8 \us\\
START to outcome, 95th, 99th percentile & 61.9, 75.2 \us\\
START to outcome, maximum & 441.4 \us\\
12 writes, 51 writes & 0.23, 0.89 ms\\
\addlinespace
Board power (whole board, external meter) & 5.1 W\\
\bottomrule
\multicolumn{2}{@{}p{0.97\columnwidth}@{}}{\footnotesize The device has
53,200 LUTs, 106,400 FFs, 140 BRAMs, and 220 DSP
blocks~\cite{xilinx2018ds190}. Compiled-driver times are medians over
2,000 sealed jobs and 100,000 repetitions of the write sequences, and
each includes about 1~\us{} for reading the clock. START-to-outcome
times in the Python runtime are over the 21,000 approved jobs. In the
gated runs logged on one board, the accelerator's run time after
release had medians of 0.72~ms and 2.09~ms for the two job types (5,700
and 1,890 jobs).}
\end{tabular}
\end{table}

\section{Evaluation}
\label{sec:eval}

Job-atomic release (Section~\ref{sec:model}) makes four claims that can be
checked on a running system: an approved job is released, a changed job
is not, an approval works only once, and a rejection disturbs nothing
else. This section first shows what a check on each write lets through,
then takes the four claims in turn, extends them to all inputs by
proof, and ends with the cost. The single bus path was shown by
construction in Section~\ref{sec:impl}.

What mixing does under a check on each write was examined in a software
model with the same cipher, the same job format, and 1,000 pairs of
approved jobs with random addresses and sizes. The software delivered the first writes of one approved
job and the remaining writes of another, at each of the 11 possible
split points. With a tag on each write, with or without a forward-only
counter, every write was accepted, and in 8,000 of the 11,000 attempts
the model accelerator started with a configuration that no authority
had approved; in the other 3,000, at the last three split points, the
mixture equals an approved job, because the control writes are the same
in every job. A counter that must be consecutive stopped the start, but
the first writes were already in the registers. Under the whole-job
rule of Algorithm~\ref{alg:dispatch}, no write was released in any
attempt. The outcome under each rule is a property of the rule and not
of where it is built: logic that checks each write and passes it on
leaves the same values in the registers as the model does. As published,
ShEF authenticates each register message and describes no counter for
them~\cite{zhao2022shef}, which is the first rule. Salus decrypts,
verifies, and forwards each register transaction under a session
counter~\cite{zou2024salus}, which is the second or the third rule,
since the paper does not say whether the counter must be consecutive.
The model applies these rules as the papers describe them and does not
reproduce their implementations. The model is the general case, in
which approved jobs differ. In the tested system every approved job of
a board carries the same three addresses, so a mixture of two approved
jobs can differ from an approved job only in its sizes, and the
hardware tests exercise altered and repeated jobs instead.

In the hardware tests, the authority ran on a separate board. The
workload was a 30-layer SmolLM2 language
model, the depth of its 135M
variant~\cite{allal2025smollm2,hf2024smollm2135m}, with fixed,
hash-checked weights and activations taken from the model. Each layer
needs four attention projections and three feed-forward projections, so
one run of the model sends 210 jobs through the gate. Jobs were
submitted one at a time and in order, and the software waited for the
accelerator's done interrupt before it submitted the next one. After
each job, a second sealed packet of one record cleared that interrupt,
so this maintenance write also went through the gate.

The verdict on every attack was read from the gate's hardware status
register and not from the application's return codes. A job counts as
released when that register shows \sig{DONE}, which the gate sets only
after both checks have passed and the accelerator has acknowledged the
last write (line~\ref{alg:done} of Algorithm~\ref{alg:dispatch}).
Release is thus the gate's statement that it issued the job's writes,
not a statement that the computation has finished. A rejection shows \sig{DONE} clear together with the authentication-fail
or the freshness-fail flag. The accelerator's registers cannot be read
back from the processor, so the absence of writes after a rejection
rests on these flags, on the structure of Fig.~\ref{fig:fsm}, and on
the proof reported below; two of the tests add a check outside the
gate.

An approved job is released. The model was run 100 times, which sent
21,000 approved jobs through the gate, and every one was verified and
released. In the final gated
runs logged on one board, the hash of the accelerator's output was
compared with a software reference for 1,980 jobs and was equal in
every case.

A changed job is not released, and an approval works only once. Two
attacks stand behind these claims (Table~\ref{tab:attacks}). In the
first, a sealed job was altered in one place and presented under
the authorization issued for the original job. The altered part was the
tag itself or the bytes of the sealed job that carry the input address,
the weight address, the output address, the sizes, or the control
settings. Each kind of change was applied to ten samples at each of the
210 job positions in the model, and none of the altered jobs was
released. In the second, jobs that had already been released were sent
again, bit for bit. These jobs still carried a valid tag, so they
passed the tag check as expected, and every one was then rejected by
the sequence check.

\begin{table}[!t]
\centering
\caption{Attack results.}
\label{tab:attacks}
\small
\setlength{\tabcolsep}{3.5pt}
\begin{tabular}{@{}lrrl@{}}
\toprule
\textbf{Attack} & \textbf{Attempts} & \textbf{Released} & \textbf{Stopped by}\\
\midrule
Changed tag              & 2,100 & 0 & tag check\\
Changed input address    & 2,100 & 0 & tag check\\
Changed weight address   & 2,100 & 0 & tag check\\
Changed output address   & 2,100 & 0 & tag check\\
Changed sizes            & 2,100 & 0 & tag check\\
Changed control settings & 2,100 & 0 & tag check\\
Repeated job             & 2,100 & 0 & sequence check\\
\midrule
All attacks              & 14,700 & 0 & \\
\bottomrule
\multicolumn{4}{@{}p{0.97\columnwidth}@{}}{\footnotesize One-sided 95\%
bound on the chance of a release per attempt, under an
independent-trial model: $2.4\times10^{-4}$ for the
12,600 altered jobs and $1.4\times10^{-3}$ for the 2,100 repeated
jobs.}
\end{tabular}
\end{table}

An approval does not survive a restart. The gate was
restarted with a new key and epoch, and a job captured before the
restart was presented. The gate rejected it at the tag check, with the
authentication-fail flag set and \sig{DONE} clear. A newly issued job
was then accepted, and the accelerator's output had the same hash as
before the restart, which checks the result outside the gate.

A rejection disturbs nothing else. Stopping an attack by halting the
system would be of little use in an embedded product, so each of the
14,700 rejected attempts was followed immediately by a fresh approved
job. Every one of these jobs was
released, with no reboot and no reconfiguration of the FPGA.

The last test looked beyond the gate. In the full system, the outputs
of the protected accelerator are consumed by a model-adaptation
workflow, in which a trainer updates the model's weights. A further
5,880 rejected attempts were issued while the gate was coupled to this
live workflow. The stored state of the trainer was monitored across the
rejected intervals: the model and weight hashes, the version and update
counters, and the training logs. None of it changed. A fresh approved
job was then released, and training advanced normally.

Table~\ref{tab:attacks} also gives, for each check, the Clopper--Pearson
bound on the chance that a single attempt is
released~\cite{clopper1934use}. The bound assumes independent trials,
and the attempts repeat positions, samples, and kinds of change, so it
is a descriptive figure under that model and not the success
probability of a real attacker. The campaign was built for coverage,
with every kind of change at every job position, and not as a
statistical test of the cipher, whose strength rests on the analysis of
Ascon~\cite{dobraunig2021ascon,turan2025ascon}. It covers the logic
around the cipher, and in no trial did the status register report a
release. The two attacks exercised the tag check and the sequence
check, which guard the release path. Shortened or reordered jobs, a
sequence number that skips ahead, writes to the gate while it is busy,
and the bus-error and configuration-error paths of Fig.~\ref{fig:fsm}
were not exercised in the campaign; the formal check that follows
covers them.

The same claims hold for inputs that no test can produce on demand,
such as a write that arrives in the few microseconds while the gate is
busy or a bus error from the accelerator. The release path was
therefore also verified formally on the gate's Verilog, a practice that AKER follows for its
wrappers~\cite{restuccia2021aker}. The cipher core was replaced by a
black box whose outputs are unconstrained, and every bus input was left
free, so the result holds for all input sequences and for any behavior
of the core. Table~\ref{tab:formal} lists the properties. There are 33,
with a further 72 checks that cover each word of the buffers. They were
proven by induction with the open-source SymbiYosys
flow~\cite{wolf2026yosys,yosyshq2026sby} and the Z3
solver~\cite{demoura2008z3} in about 15 seconds. The proof shows that
the gate never issues a write unless the core has reported a valid tag;
that the core computes the cipher correctly rests on its agreement with
the NIST test vectors (Section~\ref{sec:impl}) and on the experiments.
The properties were themselves checked in two ways. Twelve faults were
seeded into the Verilog one at a time, among them a release without a
valid tag, a missing sequence comparison, a sequence register updated
after dispatch, and a staged register that can change while a job is in
progress, and the proof failed for every one. A bounded search from
reset also reached all six outcomes of Fig.~\ref{fig:fsm}, so the
properties do not hold vacuously. The proof is on the Verilog source,
with reset asserted once at the start, and not on the routed netlist.

\begin{table*}[!t]
\centering
\caption{Properties of the gate proven on its Verilog for all input
sequences.}
\label{tab:formal}
\small
\setlength{\tabcolsep}{4pt}
\begin{tabular}{@{}r L{0.94\textwidth}@{}}
\toprule
 & \textbf{Property}\\
\midrule
1 & A write appears on the accelerator's bus only in the state that sends it, and the address and data on the bus are the quarantined record at the current index.\\
2 & Dispatch is entered only after the cipher core reports a valid tag, with the sequence number equal to the nonce's low half and higher than every number accepted before.\\
3 & The sequence register only moves forward, changes only at such a grant, and already holds the job's number in the cycle in which dispatch begins.\\
4 & Each granted job has a sequence number above that of the job granted before it, so no job is granted twice.\\
5 & While a job is in progress, and in the cycle in which the start request is taken, the key, nonce, lengths, tag, sequence word, and sealed records cannot change.\\
6 & The quarantined records cannot change between the grant and the end of dispatch; they are sent in order, one per acknowledged write, and their number equals the declared length.\\
7 & A job that fails the tag or the sequence check returns the machine to \sig{IDLE} with no write. A bus error from the accelerator ends the job at that write, after the earlier writes have been issued, and \sig{DONE} is set only after the last record has been acknowledged without error.\\
8 & A job is taken up only when the key is locked and the request is well formed; the lock is one-way, a locked key cannot change, and the key reads back as zero.\\
9 & The gate never issues a read on the accelerator's bus.\\
\bottomrule
\end{tabular}
\end{table*}

The property costs about 10~\us{} per job
(Table~\ref{tab:impl}). A compiled driver that writes the registers
directly was timed on one board with 2,000 sealed jobs under a test
key, whose records wrote only argument registers. A sealed job takes
about 13~\us{} in all, against about 3~\us{} for a direct start of 12
writes over the same port. Both times start with the job already in the
processor's memory; the authority's sealing of a job and its delivery
to the processor are not included. Most of that time goes to staging the job;
the gate itself needs under 4~\us{} to decrypt it, check the tag and
the sequence number, and issue the writes. The gate handles the same
number of records for every job, whatever the size of the matrices, and
data transfers do not pass through it, so this cost is fixed. Beside
the accelerator's run time after release in the gated runs logged on
one board, it is between about 0.5\% and 1.4\% of a job.

The campaign itself ran on a Python runtime, in which a single register
write takes longer than the gate's whole check. The times logged there,
from the \sig{START} write to the outcome and without staging,
therefore show mostly the interpreter, since the gate had finished
before the first status read. The cause of the rare slowest case was
not traced, and scheduling of the software thread is one possible
explanation. The sealed packet that clears the accelerator's interrupt
adds a smaller cost of the same kind, which was not timed separately.
The direct start was timed over the same bus port and not on a build
without the gate. Whether this cost is small can be judged only beside
other protections, which the next section does.

\section{Comparison with published systems}
\label{sec:compare}

\begin{figure}[!t]
\centering
\includegraphics[width=\columnwidth]{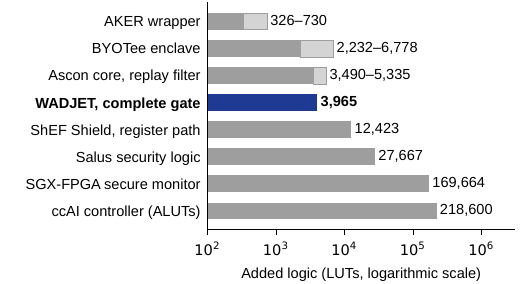}
\caption{Logic added by each FPGA design, as reported; devices,
ciphers, and protected scope differ (Table~\ref{tab:quant}).}
\label{fig:luts}
\end{figure}

Table~\ref{tab:quant} sets WADJET beside ten published systems of
Section~\ref{sec:related} that report hardware or runtime cost, using
the figures each paper gives, and Fig.~\ref{fig:luts} shows the logic
that each FPGA design adds, with a range where a paper reports several
configurations. The systems differ in what they protect, in
their platforms, and in how they report runtime cost, so the table
cannot place them on one scale. The Xilinx devices in the table all use
six-input
LUTs~\cite{xilinx2018ds190,xilinx2016ug474,amd2025ug574}, so their LUT
counts can be set side by side, as context and not as a normalized
cost; ccAI reports the adaptive LUTs of an Intel device.

\begin{table*}[!t]
\centering
\caption{Comparison of cost with published systems, as reported by each
paper.}
\label{tab:quant}
\footnotesize
\setlength{\tabcolsep}{3pt}
\begin{tabular}{@{}L{0.085\textwidth} L{0.115\textwidth} L{0.155\textwidth} L{0.115\textwidth} r r r L{0.175\textwidth}@{}}
\toprule
\textbf{System} & \textbf{Platform} & \textbf{Protects} & \textbf{Cryptography} & \textbf{LUT} & \textbf{FF} & \textbf{BRAM} & \textbf{Reported runtime cost}\\
\midrule
AKER \cite{restuccia2021aker} & Zynq UltraScale+ & single bus accesses & none & 326--730$^{a}$ & 358--744$^{a}$ & n.r. & one clock cycle per AXI transaction\\
\addlinespace
ShEF \cite{zhao2022shef} & Amazon F1 cloud FPGA (VU9P) & register and memory traffic & AES-CTR, HMAC or PMAC & 12,423$^{b}$ & 7,432$^{b}$ & 0$^{b}$ & almost none (register-only accelerator); 1.20$\times$--3.83$\times$ where memory is protected\\
\addlinespace
SGX-FPGA \cite{xia2021sgxfpga} & Virtex-7 & traffic between an SGX enclave and the FPGA & PUF-derived keys; cipher not named & 169,664$^{c}$ & 81,766$^{c}$ & 334$^{c}$ & grows with data, about 0.06 ms per KB\\
\addlinespace
Salus \cite{zou2024salus} & Alveo U200 & bitstream and register traffic & AES, HMAC, SipHash & 27,667 & 29,631 & 88 & 1.00$\times$--1.05$\times$ on the FPGA side\\
\addlinespace
ccAI \cite{wang2025ccai} & Agilex 7 & every bus packet of an unmodified accelerator & AES-GCM & 218,600$^{d}$ & 195,700 & 630 & 0.05\%--5.67\%\\
\addlinespace
Ascon core \cite{kurian2025ascon} & Artix-7 & a message stream; drives no device & Ascon-AEAD128, Bloom filter & 3,490--5,335$^{e}$ & 4,285--4,287$^{e}$ & 32 & n.r.\\
\addlinespace
BYOTee \cite{armanuzzaman2024byotee} & Zynq-7000 (Cora Z7-07S) & computation moved into soft-processor enclaves & AES-256, HMAC, BLAKE2 & 2,232--6,778$^{f}$ & 2,259--7,721$^{f}$ & 16--45.5$^{f}$ & n.r.\\
\addlinespace
CAGE \cite{wang2026cage} & Arm Juno with VCU118 & tasks and data (software monitor) & HMAC task signatures & none & none & none & 9.61\%--16.30\% on FPGA workloads\\
\addlinespace
StrongBox \cite{deng2022strongbox} & Arm Juno R2 (Mali GPU) & GPU tasks and data (software monitor) & AES-128, SHA-256 & none & none & none & 4.70\%--15.26\%\\
\addlinespace
Graviton \cite{volos2018graviton} & NVIDIA GPUs (changes emulated) & GPU command groups and memory & AES-GCM & n.r. & n.r. & n.r. & 17\%--33\%\\
\addlinespace
\textbf{WADJET} & \textbf{Zynq-7020} & \textbf{whole jobs on the control path} & \textbf{Ascon-AEAD128} & \textbf{3,965} & \textbf{6,685} & \textbf{0} & \textbf{13~\us{} per sealed job against 3~\us{} for a direct start}$^{g}$\\
\bottomrule
\multicolumn{8}{@{}p{0.98\textwidth}@{}}{\footnotesize $^{a}$Per
wrapper, for 4 to 16 protected regions. $^{b}$Sum, as LUTs and FFs, of
the controller (2,348; 547), register interface (3,251; 1,902), AES-16x
(2,898; 2,347), and HMAC (3,926; 2,636) components in ShEF's Table~1.
This is the register-only Shield of its Bitcoin accelerator (1.4\% of
the device's LUTs); Shields that also protect memory take 3.3\%--11\%.
$^{c}$Secure monitor; its physical unclonable function (PUF) adds
6,190 LUTs, 9,783 FFs, and 5 BRAMs.
$^{d}$Adaptive LUTs (ALUTs). $^{e}$Smallest to largest of three
variants. $^{f}$Per enclave, over four example enclaves without
debugger. $^{g}$Compiled driver; the separate authority board is not counted. n.r.: not
reported. AES: Advanced Encryption Standard (CTR: counter mode; GCM:
Galois/counter mode); HMAC, PMAC: message authentication codes; SHA:
Secure Hash Algorithm.}
\end{tabular}
\end{table*}

In size, the complete gate has about one third of the LUTs that ShEF
needs to protect a register path and about one seventh of the LUTs of
the security logic in Salus. In flip-flops it is close to ShEF's
register path and under a quarter of Salus. These figures cover the
logic on the protected device; the authority runs on a separate board
and is not counted. Unlike Salus, it uses no
block RAM. Part of the difference is the cipher: the whole gate, Ascon
included, has fewer LUTs than the two cipher engines of a Shield alone. Both systems also do more, since ShEF
protects memory traffic and Salus the bitstream. SGX-FPGA and ccAI,
which secure a whole channel between a protected host environment and
an FPGA or accelerator, need more than forty times as many LUTs. Three designs
are smaller or similar. An AKER wrapper is five to twelve times smaller
than the gate, but it has no cryptography and keeps no state about past
jobs. The Ascon core with a replay filter is close to the gate in LUTs
and adds 32 block RAMs for its Bloom filter; it protects a message
stream, whereas the gate's figure also covers the buffers, the dispatch
unit, and both bus ports. A BYOTee enclave is of similar size and uses
16 or more block RAMs, which the accelerator of this work does not
leave free on the Zynq-7020. CAGE and StrongBox add no hardware and instead
run a software monitor in Arm CCA or TrustZone on every task; the
Cortex-A9 of a Zynq-7000 does not offer
CCA~\cite{li2022cca,xilinx2018ds190}.

In runtime cost, the systems fall into two groups. Systems that encrypt
and authenticate data as it moves pay as the data grows. ShEF reports
almost no overhead for its register-only accelerator and up to 3.83
times the run time where memory is protected, SGX-FPGA's overhead grows
with the size of the input and output, and Graviton reports 17\% to
33\%. The others report a cost per transaction or per task. AKER adds
one clock cycle to each bus transaction. Salus reports almost no
slowdown on the FPGA side with a TEE-enabled host. StrongBox reports
5\% to 15\%, and CAGE reports 10\% to 16\% overhead on FPGA workloads,
which it attributes to security operations that do not depend on the
data, such as switches between the host and its monitor. ccAI, which
checks and encrypts every packet in dedicated logic, reports under 6\%.
WADJET belongs to the second group. With a compiled driver, a sealed
job costs about 13~\us{} against about 3~\us{} for a direct start,
whatever the size of the matrices, and it leaves data transfers
untouched.

Cost has to be read against what it buys. As Table~\ref{tab:features}
showed, the FPGA designs that cost more than the gate do not verify a
job as a whole, and the systems that verify a whole task need a
software monitor on the processor or a modified device. The gate
supplies that check in FPGA logic at a cost between that of a bus
wrapper and that of a full shield.

\section{Discussion}
\label{sec:discussion}

WADJET decides whether a job may start, and it leaves two tasks to the
platform around it. The first is control of the programmable logic. On
the Zynq-7000, software on the processor can reset the logic, change
its clock, and load a different
configuration~\cite{amd2026ug585}. A product must therefore keep these
controls and the loading of the key in the processor's secure
world~\cite{xilinx2014ug1019}. It must also load a fresh key after
every reset, because a reset clears the sequence register; storing the
sequence number across power cycles can follow established
practice~\cite{parno2011memoir}. This asks less of TrustZone than a
software monitor does, since the secure world guards three controls and
one key and checks no jobs. The prototype loads and locks the key at
start-up; secure boot, key storage, and the restriction of these
controls belong to the platform and are outside the present
implementation. The second task is the data itself. The gate fixes the
addresses, the sizes, and the jobs that the accelerator works on, and a
scheme such as ShEF, or CAGE on a processor with Arm CCA, can be added
where the contents of memory must also be
protected~\cite{zhao2022shef,wang2026cage}.

The gate is not tied to this accelerator. It needs only a
register-style bus on each side and does not interpret the registers it
writes, so it can sit in front of any accelerator whose job is a list
of register writes ending in a start bit. To reuse it, the authority
produces that accelerator's register list, and the bus is wired so that
only the gate reaches the accelerator's registers. The proof of Section~\ref{sec:eval} already covers this: it leaves every register
address, value, and job length free, so it holds for every job of 1 to
32 records and not only for the 12-record job of this accelerator. In
the software model, job formats of 7 and 32 writes gave the same
outcome as the 12-write format.
Serving several
streams of jobs at once would need one sequence register per stream.

\section{Conclusion}
\label{sec:conclusion}

The fault addressed here comes from inside. The software that operates
an accelerator changes one register write or resends a job, and the
hardware runs it. The protections reviewed in this paper decide who may
reach an accelerator or keep its data safe, and none of them offers, on
a device of this class, the three things that settle whether one job
was approved and whether that approval is still unused: a single
guarded path, a whole job held back until it is checked, and a record
of approvals already used. WADJET puts the three in one gate and so
provides job-atomic release. Software
submits, a separate authority approves, and the gate in the FPGA logic,
the processor's only bus route to an unmodified accelerator's
registers, uses up each approval as it releases the job, writing the
registers itself with the start bit last. The processor gives up direct
control of the accelerator's registers and still submits approved work,
through a boundary that the logic enforces.

The fix holds, and its cost is small. In a software model, a check on
each write ran an unapproved job in 8,000 of 11,000 mixing attempts,
and the whole-job rule released no write. The release path is proven on the
gate's Verilog for all input sequences. On a Zynq-7020 serving a
language model, the gate released all 21,000 approved jobs,
none of 14,700 altered or repeated jobs, and a fresh job after every
rejection. It needs no block RAM beside an accelerator that fills the
device's block RAM, and a sealed job costs about ten microseconds more than a direct
start, a cost that does not grow with the data. Among the compared systems, WADJET is the only one that
verifies a whole job in FPGA logic, in front of an unmodified
accelerator, and releases it at most once, with about one third of the
LUTs that ShEF needs to protect a register path and about one seventh
of those of the security logic in Salus.

The gate is the piece that those systems lack and not a replacement for
them. ShEF and Salus also protect data or the bitstream, and the gate
works beside such protection. Its own guarantee rests on trusted
loading of the key and on untrusted software being unable to reset or
reconfigure the logic. Future work will measure the cost on a build
without the gate, release each job only when the accelerator is idle,
and serve several streams of jobs.

\section*{CRediT authorship contribution statement}
\textbf{Mohamed El-Hadedy:} Conceptualization, Methodology, Software,
Investigation, Validation, Formal analysis, Writing -- original draft,
Writing -- review \& editing, Funding acquisition.

\section*{Declaration of competing interest}
The author declares no known competing financial interests or personal
relationships that could have appeared to influence the work reported
in this paper.

\section*{Data availability}
The measurement logs that support this study are available from the
corresponding author on request. The register map and the job format
needed to rebuild the gate are given in Sections~\ref{sec:design}
and~\ref{sec:impl}. The formal properties and the timing programs are
available on the same terms. The hardware design files are not publicly
available because they are part of an ongoing sponsored project.

\section*{Acknowledgments}
This work was supported by the U.S. Army Research Office (ARO) under
Grant W911NF-24-1-0265, the U.S. Navy Naval Engineering Education
Consortium (NEEC) under Grant N001742310002, the Office of Naval
Research (ONR) Summer Faculty Research Program, and the Air Force
Research Laboratory (AFRL) under agreement FA8650-24-2-2403. The U.S.
Government is authorized to reproduce and distribute reprints for
Government purposes, notwithstanding any copyright notation thereon;
the views and conclusions are those of the author and do not
necessarily reflect the official policies or endorsements of ARO, ONR,
AFRL, or the U.S. Government. The sponsors had no role in the design of
the study, in the collection and analysis of the data, or in the
writing of this article.

\section*{Declaration of generative AI and AI-assisted technologies in
the manuscript preparation process}
During the preparation of this work, the author used generative AI tools
to assist with manuscript drafting, language revision, organization,
figure preparation, and consistency checks across the technical
descriptions, figures, and experimental results. The figures illustrate
the author's proposed architecture and experimental findings, together
with comparisons based on cited published work; AI tools assisted with
their drawing and visual refinement. The author conceived the research
idea, designed the architecture and protocol, and conducted the
experiments. After using these tools, the author reviewed and edited the
content as needed and retains full responsibility for the research
design, implementation, scientific interpretation, validation of
results, and the content of the published article.

\bibliographystyle{elsarticle-num}
\bibliography{refs}

\begin{thebibliography}{10}
\expandafter\ifx\csname url\endcsname\relax
  \def\url#1{\texttt{#1}}\fi
\expandafter\ifx\csname urlprefix\endcsname\relax\def\urlprefix{URL }\fi
\expandafter\ifx\csname href\endcsname\relax
  \def\href#1#2{#2} \def\path#1{#1}\fi

\bibitem{sze2017efficient}
V.~Sze, Y.-H. Chen, T.-J. Yang, J.~S. Emer, Efficient processing of deep neural
  networks: A tutorial and survey, Proceedings of the IEEE 105~(12) (2017)
  2295--2329.
\newblock \href {https://doi.org/10.1109/JPROC.2017.2761740}
  {\path{doi:10.1109/JPROC.2017.2761740}}.

\bibitem{xu2024llamaf}
H.~Xu, Y.~Li, S.~Ji, {LlamaF}: An efficient {Llama2} architecture accelerator
  on embedded {FPGAs} (2024).
\newblock \href {http://arxiv.org/abs/2409.11424} {\path{arXiv:2409.11424}}.

\bibitem{huggingface2026intrusion}
{Hugging Face},
  \href{https://huggingface.co/blog/agent-intrusion-technical-timeline}{Anatomy
  of a frontier lab agent intrusion: A technical timeline of the {July} 2026
  incident}, Hugging Face Blog, accessed 14 November 2026 (27~Jul. 2026).
\newline\urlprefix\url{https://huggingface.co/blog/agent-intrusion-technical-timeline}

\bibitem{openai2026hfevalincident}
{OpenAI},
  \href{https://openai.com/index/hugging-face-model-evaluation-security-incident/}{{OpenAI}
  and {Hugging Face} partner to address security incident during model
  evaluation}, accessed 14 November 2026 (21~Jul. 2026).
\newline\urlprefix\url{https://openai.com/index/hugging-face-model-evaluation-security-incident/}

\bibitem{pinto2019trustzone}
S.~Pinto, N.~Santos, Demystifying {Arm} {TrustZone}: A comprehensive survey,
  ACM Computing Surveys 51~(6) (2019) 130:1--130:36.
\newblock \href {https://doi.org/10.1145/3291047} {\path{doi:10.1145/3291047}}.

\bibitem{benhani2019trustzone}
E.~M. Benhani, L.~Bossuet, A.~Aubert, The security of {ARM} {TrustZone} in a
  {FPGA}-based {SoC}, IEEE Transactions on Computers 68~(8) (2019) 1238--1248.
\newblock \href {https://doi.org/10.1109/TC.2019.2900235}
  {\path{doi:10.1109/TC.2019.2900235}}.

\bibitem{turan2025ascon}
M.~{S{\"o}nmez Turan}, K.~A. McKay, D.~Chang, J.~Kang, J.~Kelsey,
  \href{https://doi.org/10.6028/NIST.SP.800-232}{Ascon-based lightweight
  cryptography standards for constrained devices: Authenticated encryption,
  hash, and extendable output functions}, NIST Special Publication 800-232,
  National Institute of Standards and Technology, Gaithersburg, MD, USA (Aug.
  2025).
\newblock \href {https://doi.org/10.6028/NIST.SP.800-232}
  {\path{doi:10.6028/NIST.SP.800-232}}.
\newline\urlprefix\url{https://doi.org/10.6028/NIST.SP.800-232}

\bibitem{amd2026ug585}
{AMD}, \href{https://docs.amd.com/r/en-US/ug585-zynq-7000-SoC-TRM}{Zynq 7000
  {SoC} technical reference manual}, UG585 (v1.15) (Feb. 2026).
\newline\urlprefix\url{https://docs.amd.com/r/en-US/ug585-zynq-7000-SoC-TRM}

\bibitem{amd2022ug1399}
{AMD}, \href{https://docs.amd.com/r/2022.2-English/ug1399-vitis-hls}{Vitis
  high-level synthesis user guide}, UG1399 (v2022.2) (Dec. 2022).
\newline\urlprefix\url{https://docs.amd.com/r/2022.2-English/ug1399-vitis-hls}

\bibitem{arm2021axi}
{Arm Limited}, {AMBA} {AXI} and {ACE} protocol specification, ARM IHI 0022H.c
  (Jan. 2021).

\bibitem{wang2026sok}
C.~Wang, J.~Huang, Y.~Liang, X.~Peng, Y.~Zhang, F.~Zhang, J.~Cao, H.~Lu,
  R.~Hou, S.~Yan, T.~Wei, Z.~He, {SoK}: Analysis of accelerator {TEE} designs,
  in: Network and Distributed System Security (NDSS) Symposium 2026, Internet
  Society, San Diego, CA, USA, 2026.
\newblock \href {https://doi.org/10.14722/ndss.2026.241424}
  {\path{doi:10.14722/ndss.2026.241424}}.

\bibitem{perkins2025sok}
G.~Perkins, B.~Macht, L.~Ritzdorf, T.~{Running Crane}, B.~LaMeres, C.~Izurieta,
  A.~M. Reinhold, {SoK}: Trusted execution in {SoC-FPGAs}, in: 2025
  Intermountain Engineering, Technology and Computing (IETC), Orem, UT, USA,
  2025.
\newblock \href {https://doi.org/10.1109/IETC64455.2025.11039468}
  {\path{doi:10.1109/IETC64455.2025.11039468}}.

\bibitem{restuccia2021aker}
F.~Restuccia, A.~Meza, R.~Kastner, {AKER}: A design and verification framework
  for safe and secure {SoC} access control, in: 2021 IEEE/ACM International
  Conference on Computer Aided Design (ICCAD), 2021, pp. 1--9.
\newblock \href {https://doi.org/10.1109/ICCAD51958.2021.9643538}
  {\path{doi:10.1109/ICCAD51958.2021.9643538}}.

\bibitem{cotret2016firewalls}
P.~Cotret, G.~Gogniat, M.~J. {Sep{\'u}lveda Fl{\'o}rez}, Protection of
  heterogeneous architectures on {FPGAs}: An approach based on hardware
  firewalls, Microprocessors and Microsystems 42 (2016) 127--141.
\newblock \href {https://doi.org/10.1016/j.micpro.2016.01.013}
  {\path{doi:10.1016/j.micpro.2016.01.013}}.

\bibitem{pinto2021selfsecured}
S.~Pinto, P.~Machado, D.~Oliveira, D.~Cerdeira, T.~Gomes, Self-secured devices:
  High performance and secure {I/O} access in {TrustZone}-based systems,
  Journal of Systems Architecture 119 (2021) 102238.
\newblock \href {https://doi.org/10.1016/j.sysarc.2021.102238}
  {\path{doi:10.1016/j.sysarc.2021.102238}}.

\bibitem{cheng2025capchecker}
J.~Cheng, A.~T. Markettos, A.~Joannou, P.~Metzger, M.~Naylor, P.~Rugg, T.~M.
  Jones, Adaptive {CHERI} compartmentalization for heterogeneous accelerators,
  in: Proceedings of the 52nd Annual International Symposium on Computer
  Architecture (ISCA), 2025, pp. 2002--2016.
\newblock \href {https://doi.org/10.1145/3695053.3731062}
  {\path{doi:10.1145/3695053.3731062}}.

\bibitem{zhao2022shef}
M.~Zhao, M.~Gao, C.~Kozyrakis, {ShEF}: Shielded enclaves for cloud {FPGAs}, in:
  Proceedings of the 27th ACM International Conference on Architectural Support
  for Programming Languages and Operating Systems (ASPLOS), 2022, pp.
  1070--1085.
\newblock \href {https://doi.org/10.1145/3503222.3507733}
  {\path{doi:10.1145/3503222.3507733}}.

\bibitem{xia2021sgxfpga}
K.~Xia, Y.~Luo, X.~Xu, S.~Wei, {SGX-FPGA}: Trusted execution environment for
  {CPU-FPGA} heterogeneous architecture, in: 2021 58th ACM/IEEE Design
  Automation Conference (DAC), 2021, pp. 301--306.
\newblock \href {https://doi.org/10.1109/DAC18074.2021.9586207}
  {\path{doi:10.1109/DAC18074.2021.9586207}}.

\bibitem{zou2024salus}
Y.~Zou, Y.~Li, S.~Wang, L.~Su, Z.~Gu, Y.~Lu, Y.~Guan, D.~Niu, M.~Gao, Y.~Xie,
  F.~Li, Salus: A practical trusted execution environment for {CPU-FPGA}
  heterogeneous cloud platforms, in: Proceedings of the 29th ACM International
  Conference on Architectural Support for Programming Languages and Operating
  Systems, Volume 4 (ASPLOS '24), 2024, pp. 252--266.
\newblock \href {https://doi.org/10.1145/3622781.3674169}
  {\path{doi:10.1145/3622781.3674169}}.

\bibitem{wang2025ccai}
C.~Wang, D.~Tang, C.~Ci, J.~Huang, Y.~Xu, F.~Zhang, J.~Cao, J.~Song, S.~Yan,
  T.~Wei, Z.~He, {ccAI}: A compatible and confidential system for {AI}
  computing, in: Proceedings of the 58th IEEE/ACM International Symposium on
  Microarchitecture (MICRO '25), 2025, pp. 340--353.
\newblock \href {https://doi.org/10.1145/3725843.3756104}
  {\path{doi:10.1145/3725843.3756104}}.

\bibitem{li2022cca}
X.~Li, X.~Li, C.~Dall, R.~Gu, J.~Nieh, Y.~Sait, G.~Stockwell,
  \href{https://www.usenix.org/conference/osdi22/presentation/li}{Design and
  verification of the {Arm} confidential compute architecture}, in: 16th USENIX
  Symposium on Operating Systems Design and Implementation (OSDI 22), USENIX
  Association, 2022, pp. 465--484.
\newline\urlprefix\url{https://www.usenix.org/conference/osdi22/presentation/li}

\bibitem{sridhara2024acai}
S.~Sridhara, A.~Bertschi, B.~Schl{\"u}ter, M.~Kuhne, F.~Aliberti, S.~Shinde,
  \href{https://www.usenix.org/conference/usenixsecurity24/presentation/sridhara}{{ACAI}:
  Protecting accelerator execution with {Arm Confidential Computing
  Architecture}}, in: 33rd USENIX Security Symposium (USENIX Security 24),
  USENIX Association, 2024, pp. 3423--3440.
\newline\urlprefix\url{https://www.usenix.org/conference/usenixsecurity24/presentation/sridhara}

\bibitem{jang2019hix}
I.~Jang, A.~Tang, T.~Kim, S.~Sethumadhavan, J.~Huh, Heterogeneous isolated
  execution for commodity {GPUs}, in: Proceedings of the 24th International
  Conference on Architectural Support for Programming Languages and Operating
  Systems (ASPLOS), 2019, pp. 455--468.
\newblock \href {https://doi.org/10.1145/3297858.3304021}
  {\path{doi:10.1145/3297858.3304021}}.

\bibitem{shen2024tedge}
J.~Shen, Y.~Chen, W.-F. Wong, E.-C. Chang, {T-Edge}: Trusted heterogeneous edge
  computing, in: 2024 Annual Computer Security Applications Conference (ACSAC),
  2024.
\newblock \href {https://doi.org/10.1109/ACSAC63791.2024.00027}
  {\path{doi:10.1109/ACSAC63791.2024.00027}}.

\bibitem{wang2025fpgatrustzone}
S.~Wang, X.~Fan, X.~Xu, S.~Wang, L.~Ju, Z.~Zhou, {FPGA-TrustZone}: Security
  extension of {TrustZone} to {FPGA} for {SoC-FPGA} heterogeneous architecture,
  in: 2025 62nd ACM/IEEE Design Automation Conference (DAC), 2025, pp. 1--6.
\newblock \href {https://doi.org/10.1109/DAC63849.2025.11132548}
  {\path{doi:10.1109/DAC63849.2025.11132548}}.

\bibitem{wang2026cage}
C.~Wang, K.~Lu, F.~Zhang, Y.~Deng, K.~Leach, J.~Cao, Z.~Ning, S.~Yan, T.~Wei,
  Z.~He, Building confidential accelerator computing environment for {Arm CCA},
  IEEE Transactions on Dependable and Secure Computing 23~(1) (2026)
  1308--1325.
\newblock \href {https://doi.org/10.1109/TDSC.2025.3615787}
  {\path{doi:10.1109/TDSC.2025.3615787}}.

\bibitem{deng2022strongbox}
Y.~Deng, C.~Wang, S.~Yu, S.~Liu, Z.~Ning, K.~Leach, J.~Li, S.~Yan, Z.~He,
  J.~Cao, F.~Zhang, {StrongBox}: A {GPU} {TEE} on {Arm} endpoints, in:
  Proceedings of the 2022 ACM SIGSAC Conference on Computer and Communications
  Security (CCS), 2022, pp. 769--783.
\newblock \href {https://doi.org/10.1145/3548606.3560627}
  {\path{doi:10.1145/3548606.3560627}}.

\bibitem{park2023grt}
H.~Park, F.~X. Lin, Safe and practical {GPU} computation in {TrustZone}, in:
  Proceedings of the Eighteenth European Conference on Computer Systems
  (EuroSys '23), 2023, pp. 505--520.
\newblock \href {https://doi.org/10.1145/3552326.3567483}
  {\path{doi:10.1145/3552326.3567483}}.

\bibitem{volos2018graviton}
S.~Volos, K.~Vaswani, R.~Bruno,
  \href{https://www.usenix.org/conference/osdi18/presentation/volos}{Graviton:
  Trusted execution environments on {GPUs}}, in: 13th USENIX Symposium on
  Operating Systems Design and Implementation (OSDI 18), USENIX Association,
  2018, pp. 681--696.
\newline\urlprefix\url{https://www.usenix.org/conference/osdi18/presentation/volos}

\bibitem{pappachan2022patent}
P.~M. Pappachan, R.~Lal,
  \href{https://patents.google.com/patent/US11496314}{Integrity protected
  command buffer execution}, U.S. Patent 11,496,314 B2, Intel Corporation,
  filed 18 December 2019; granted 8 November 2022 (2022).
\newline\urlprefix\url{https://patents.google.com/patent/US11496314}

\bibitem{armanuzzaman2024byotee}
M.~Armanuzzaman, A.-R. Sadeghi, Z.~Zhao, Building your own trusted execution
  environments using {FPGA}, in: Proceedings of the 19th ACM Asia Conference on
  Computer and Communications Security (ASIA CCS), 2024, pp. 1584--1599.
\newblock \href {https://doi.org/10.1145/3634737.3637644}
  {\path{doi:10.1145/3634737.3637644}}.

\bibitem{hua2022guardnn}
W.~Hua, M.~Umar, Z.~Zhang, G.~E. Suh, {GuardNN}: Secure accelerator
  architecture for privacy-preserving deep learning, in: Proceedings of the
  59th ACM/IEEE Design Automation Conference (DAC), 2022, pp. 349--354.
\newblock \href {https://doi.org/10.1145/3489517.3530439}
  {\path{doi:10.1145/3489517.3530439}}.

\bibitem{yu2026lgt4cg}
F.~Yu, C.~Huang, Z.~Hua, Y.~Ding, K.~Wang, R.~Yang, H.~Cheng, B.~Yu, {LGT4CG}:
  Lightweight {GPU-TEE} for cloud {GPUs}, Journal of Systems Architecture 178
  (2026) 103886.
\newblock \href {https://doi.org/10.1016/j.sysarc.2026.103886}
  {\path{doi:10.1016/j.sysarc.2026.103886}}.

\bibitem{zhu2020hetee}
J.~Zhu, R.~Hou, X.~Wang, W.~Wang, J.~Cao, B.~Zhao, Z.~Wang, Y.~Zhang, J.~Ying,
  L.~Zhang, D.~Meng, Enabling rack-scale confidential computing using
  heterogeneous trusted execution environment, in: 2020 IEEE Symposium on
  Security and Privacy (SP), 2020, pp. 1450--1465.
\newblock \href {https://doi.org/10.1109/SP40000.2020.00054}
  {\path{doi:10.1109/SP40000.2020.00054}}.

\bibitem{schneider2021composite}
M.~Schneider, A.~Dhar, I.~Puddu, K.~Kostiainen, S.~{\v{C}}apkun, Composite
  enclaves: Towards disaggregated trusted execution, IACR Transactions on
  Cryptographic Hardware and Embedded Systems 2022~(1) (2022) 630--656.
\newblock \href {https://doi.org/10.46586/tches.v2022.i1.630-656}
  {\path{doi:10.46586/tches.v2022.i1.630-656}}.

\bibitem{tian2014gvirt}
K.~Tian, Y.~Dong, D.~Cowperthwaite,
  \href{https://www.usenix.org/conference/atc14/technical-sessions/presentation/tian}{A
  full {GPU} virtualization solution with mediated pass-through}, in: 2014
  USENIX Annual Technical Conference (USENIX ATC 14), USENIX Association, 2014,
  pp. 121--132.
\newline\urlprefix\url{https://www.usenix.org/conference/atc14/technical-sessions/presentation/tian}

\bibitem{mai2023honeycomb}
H.~Mai, J.~Zhao, H.~Zheng, Y.~Zhao, Z.~Liu, M.~Gao, C.~Wang, H.~Cui, X.~Feng,
  C.~Kozyrakis,
  \href{https://www.usenix.org/conference/osdi23/presentation/mai}{Honeycomb:
  Secure and efficient {GPU} executions via static validation}, in: 17th USENIX
  Symposium on Operating Systems Design and Implementation (OSDI 23), USENIX
  Association, 2023, pp. 155--172.
\newline\urlprefix\url{https://www.usenix.org/conference/osdi23/presentation/mai}

\bibitem{markettos2019thunderclap}
A.~T. Markettos, C.~Rothwell, B.~F. Gutstein, A.~Pearce, P.~G. Neumann, S.~W.
  Moore, R.~N.~M. Watson, Thunderclap: Exploring vulnerabilities in operating
  system {IOMMU} protection via {DMA} from untrustworthy peripherals, in:
  Network and Distributed Systems Security (NDSS) Symposium 2019, Internet
  Society, San Diego, CA, USA, 2019.
\newblock \href {https://doi.org/10.14722/ndss.2019.23194}
  {\path{doi:10.14722/ndss.2019.23194}}.

\bibitem{feng2024siopmp}
E.~Feng, D.~Feng, D.~Du, Y.~Xia, W.~Zheng, S.~Zhao, H.~Chen, {sIOPMP}: Scalable
  and efficient {I/O} protection for {TEEs}, in: Proceedings of the 29th ACM
  International Conference on Architectural Support for Programming Languages
  and Operating Systems, Volume 2 (ASPLOS '24), 2024, pp. 1061--1076.
\newblock \href {https://doi.org/10.1145/3620665.3640378}
  {\path{doi:10.1145/3620665.3640378}}.

\bibitem{parno2011memoir}
B.~Parno, J.~R. Lorch, J.~R. Douceur, J.~Mickens, J.~M. McCune, Memoir:
  Practical state continuity for protected modules, in: 2011 IEEE Symposium on
  Security and Privacy (SP), 2011, pp. 379--394.
\newblock \href {https://doi.org/10.1109/SP.2011.38}
  {\path{doi:10.1109/SP.2011.38}}.

\bibitem{kurian2025ascon}
M.~G. Kurian, Y.~Chen, Ascon on {FPGA}: Post-quantum safe authenticated
  encryption with replay protection for {IoT}, Electronics 14~(13) (2025) 2668.
\newblock \href {https://doi.org/10.3390/electronics14132668}
  {\path{doi:10.3390/electronics14132668}}.

\bibitem{dobraunig2021ascon}
C.~Dobraunig, M.~Eichlseder, F.~Mendel, M.~Schl{\"a}ffer, Ascon v1.2:
  Lightweight authenticated encryption and hashing, Journal of Cryptology
  34~(3) (2021) 33.
\newblock \href {https://doi.org/10.1007/s00145-021-09398-9}
  {\path{doi:10.1007/s00145-021-09398-9}}.

\bibitem{gross2017ascon}
H.~Gro{\ss}, E.~Wenger, C.~Dobraunig, C.~Ehrenh{\"o}fer, Ascon hardware
  implementations and side-channel evaluation, Microprocessors and Microsystems
  52 (2017) 470--479.
\newblock \href {https://doi.org/10.1016/j.micpro.2016.10.006}
  {\path{doi:10.1016/j.micpro.2016.10.006}}.

\bibitem{andreeva2014rup}
E.~Andreeva, A.~Bogdanov, A.~Luykx, B.~Mennink, N.~Mouha, K.~Yasuda, How to
  securely release unverified plaintext in authenticated encryption, in:
  Advances in Cryptology -- ASIACRYPT 2014, Part I, Vol. 8873 of Lecture Notes
  in Computer Science, Springer, 2014, pp. 105--125.
\newblock \href {https://doi.org/10.1007/978-3-662-45611-8_6}
  {\path{doi:10.1007/978-3-662-45611-8_6}}.

\bibitem{kent2005rfc4303}
S.~Kent, {IP} encapsulating security payload ({ESP}), RFC 4303, Internet
  Engineering Task Force (Dec. 2005).
\newblock \href {https://doi.org/10.17487/RFC4303}
  {\path{doi:10.17487/RFC4303}}.

\bibitem{xilinx2018ds190}
{Xilinx},
  \href{https://docs.amd.com/v/u/en-US/ds190-Zynq-7000-Overview}{Zynq-7000
  {SoC} data sheet: Overview}, DS190 (v1.11.1) (Jul. 2018).
\newline\urlprefix\url{https://docs.amd.com/v/u/en-US/ds190-Zynq-7000-Overview}

\bibitem{tul2019pynqz2}
{TUL}, \href{https://www.tulembedded.com/FPGA/ProductsPYNQ-Z2.html}{{PYNQ-Z2}
  reference manual}, v1.1 (Oct. 2019).
\newline\urlprefix\url{https://www.tulembedded.com/FPGA/ProductsPYNQ-Z2.html}

\bibitem{amd2026pg247}
{AMD}, \href{https://docs.amd.com/r/en-US/pg247-smartconnect}{{SmartConnect}
  {LogiCORE} {IP} product guide}, PG247 (v1.0) (Jun. 2026).
\newline\urlprefix\url{https://docs.amd.com/r/en-US/pg247-smartconnect}

\bibitem{allal2025smollm2}
L.~{Ben Allal}, A.~Lozhkov, E.~Bakouch, G.~{Mart{\'\i}n Bl{\'a}zquez},
  G.~Penedo, L.~Tunstall, A.~Marafioti, H.~Kydl{\'\i}{\v{c}}ek, A.~{Piqueres
  Lajar{\'\i}n}, V.~Srivastav, J.~Lochner, C.~Fahlgren, X.-S. Nguyen,
  C.~Fourrier, B.~Burtenshaw, H.~Larcher, H.~Zhao, C.~Zakka, M.~Morlon,
  C.~Raffel, L.~von Werra, T.~Wolf, {SmolLM2}: When smol goes big ---
  data-centric training of a small language model (2025).
\newblock \href {http://arxiv.org/abs/2502.02737} {\path{arXiv:2502.02737}},
  \href {https://doi.org/10.48550/arXiv.2502.02737}
  {\path{doi:10.48550/arXiv.2502.02737}}.

\bibitem{hf2024smollm2135m}
{Hugging Face},
  \href{https://huggingface.co/HuggingFaceTB/SmolLM2-135M}{{SmolLM2-135M}},
  Model repository, config.json, accessed 14 November 2026 (2024).
\newline\urlprefix\url{https://huggingface.co/HuggingFaceTB/SmolLM2-135M}

\bibitem{clopper1934use}
C.~J. Clopper, E.~S. Pearson, The use of confidence or fiducial limits
  illustrated in the case of the binomial, Biometrika 26~(4) (1934) 404--413.
\newblock \href {https://doi.org/10.1093/biomet/26.4.404}
  {\path{doi:10.1093/biomet/26.4.404}}.

\bibitem{wolf2026yosys}
C.~Wolf, \href{https://yosyshq.net/yosys/}{Yosys open {SYnthesis} suite},
  Version 0.71, accessed 14 November 2026 (2026).
\newline\urlprefix\url{https://yosyshq.net/yosys/}

\bibitem{yosyshq2026sby}
{YosysHQ}, \href{https://yosyshq.readthedocs.io/projects/sby/}{{SymbiYosys}
  (sby): Front-end for {Yosys}-based formal verification flows}, accessed 14
  November 2026 (2026).
\newline\urlprefix\url{https://yosyshq.readthedocs.io/projects/sby/}

\bibitem{demoura2008z3}
L.~de~Moura, N.~Bj{\o}rner, {Z3}: An efficient {SMT} solver, in: Tools and
  Algorithms for the Construction and Analysis of Systems (TACAS 2008), Vol.
  4963 of Lecture Notes in Computer Science, Springer, 2008, pp. 337--340.
\newblock \href {https://doi.org/10.1007/978-3-540-78800-3_24}
  {\path{doi:10.1007/978-3-540-78800-3_24}}.

\bibitem{xilinx2016ug474}
{Xilinx}, \href{https://docs.amd.com/v/u/en-US/ug474_7Series_CLB}{7 series
  {FPGAs} configurable logic block user guide}, UG474 (v1.8) (Sep. 2016).
\newline\urlprefix\url{https://docs.amd.com/v/u/en-US/ug474_7Series_CLB}

\bibitem{amd2025ug574}
{AMD}, \href{https://docs.amd.com/r/en-US/ug574-ultrascale-clb}{{UltraScale}
  architecture configurable logic block user guide}, UG574 (v1.6) (Jan. 2025).
\newline\urlprefix\url{https://docs.amd.com/r/en-US/ug574-ultrascale-clb}

\bibitem{xilinx2014ug1019}
{Xilinx},
  \href{https://docs.amd.com/v/u/en-US/ug1019-zynq-trustzone}{Programming {ARM}
  {TrustZone} architecture on the {Xilinx} {Zynq-7000} all programmable {SoC}},
  UG1019 (v1.0) (May 2014).
\newline\urlprefix\url{https://docs.amd.com/v/u/en-US/ug1019-zynq-trustzone}

\end{thebibliography}

\end{document}